\documentclass[aps,A4,prl, superscriptaddress ,preprintnumbers,amsmath,amssymb,floatfix,footinbib,twocolumn]{revtex4}

\usepackage{graphicx}
\usepackage{epstopdf}
\usepackage{amsmath}
\usepackage{graphicx,color}

\usepackage{soul}

\newcommand{\av}[1]{\left\langle{#1}\right\rangle}

\newcommand{\ket}[1]{\vert#1\rangle}

\newcommand{\nn}{\\\nonumber}
\newcommand{\half}{\frac{1}{2}}
\newcommand{\snr}{{\rm SNR}}

\begin{document}

\title{Non-absorbing high-efficiency counter for itinerant microwave photons}


\author{Bixuan Fan}
\affiliation{Center for Engineered Quantum Systems, School of Mathematics and
Physics, The University of Queensland, St Lucia, Queensland 4072, Australia}

\author{G\"{o}ran Johansson}
\affiliation{Microtechnology and Nanoscience, Chalmers University of Technology,
S-41296, G\"{o}teborg, Sweden}

\author{Joshua Combes}
\affiliation{Center for Quantum Information and Control, University of New Mexico,
Albuquerque, NM 87131-0001, USA. }

\author{G. J. Milburn}
\affiliation{Center for Engineered Quantum Systems, School of Mathematics and
Physics, The University of Queensland, St Lucia, Queensland 4072, Australia}

\author{Thomas M. Stace}\email[]{stace@physics.uq.edu.au}
\affiliation{Center for Engineered Quantum Systems, School of Mathematics and
Physics, The University of Queensland, St Lucia, Queensland 4072, Australia}

\begin{abstract}
Detecting an itinerant microwave photon with high efficiency is an outstanding problem in microwave photonics and its applications.  We present a scheme to detect an itinerant microwave photon in a transmission line via the nonlinearity provided by a transmon in a driven microwave resonator. 
{With a single transmon we achieve 84\% distinguishability between zero and one microwave photons and 90\% distinguishability with two cascaded transmons by performing continuous measurements on the output field of the resonator.}   We also show how the measurement diminishes coherence in the photon number basis thereby illustrating a fundamental principle of quantum measurement: the higher the measurement efficiency, the faster is the decoherence.
\end{abstract}

\maketitle

Since the early theoretical work on photodetection \cite{PhysRev.130.2529,mandel1964theory} both  theory and technology have advanced dramatically. Conventional photon detectors, such as avalanche photodiode (APD) and photomultiplier tube (PMT), are widely used in practice. However, they destroy the {signal} photon {during} detection. There are a number of schemes for quantum non-demolition (QND) optical photon detection \cite{PhysRevA.71.033819,PhysRevA.79.052115,reiserer2013nondestructive} but typically they require a high-Q cavity for storing the signal mode containing the photon(s) to be detected, and a leaky cavity for manipulating and detecting the probe mode. Thus, during one lifetime of a signal photon, the probe mode undergoes many cycles to accumulate information about the signal. This type of detection requires repeated measurements and the high-Q cavity limits the  photodetection bandwidth. In the microwave regime the detection of single photons \cite{gleyzes2007quantum,romero2009photodetection,PhysRevLett.102.173602,johnson2010quantum,PhysRevLett.107.217401,PhysRevA.84.063834,PhysRevA.86.032311,PhysRevB.86.174506,PhysRevLett.112.080401,PhysRevLett.112.093601} is more challenging, especially non-destructive detection \cite{gleyzes2007quantum,johnson2010quantum,PhysRevLett.112.080401,PhysRevLett.112.093601}. Here we propose a scheme for non-absorbing, high-efficiency detection of single itinerant microwave photons via the nonlinearity provided by an artificial superconducting atom, a transmon \cite{PhysRevA.76.042319}.

Previously \cite{PhysRevLett.110.053601,PhysRevLett.112.093601}, we considered schemes where the signal photon wave packet propagates freely in a open transmission line \cite{PhysRevA.84.063834,PhysRevLett.111.053601} and encounters the lowest transition of a transmon. The cw-probe field couples the first and second excited states of the transmon and is monitored via continuous homodyne detection. Displacements in the homodyne current, due to the large transmon-induced cross-Kerr non-linearity \cite{PhysRevLett.111.053601}, indicate the presence of a photon.  We showed that, in spite of the exceptionally large cross-Kerr nonlinearity it exhibits \cite{PhysRevLett.111.053601},  a single transmon in an open transmission line is insufficient for reliable microwave photon detection, due to  saturation  of the transmon response to the probe field  \cite{PhysRevLett.110.053601}.  More recently, \cite{PhysRevLett.112.093601}, we showed that multiple cascaded transmons  could achieve reliable microwave photon counting in principle, though the number of  transmons and circulators required in this scheme presents serious  {experimental challenges}.

In this Letter we propose a scheme that achieves reliable photon counting with as few as a single transmon. The key insight is to use a cavity resonant with the {\em probe field} to enhance the probe displacements, which depends on the signal photon number. For a single transmon, we report a signal-to-noise ratio (SNR) of 1.2,  corresponding to a distinguishability of $F=84\%$ between 0 and 1 photons in the signal  (i.e.\ the probability of correctly discriminating these two states). This can be improved using more transmons \cite{PhysRevLett.112.093601}, and we show that with two cascaded transmons the distinguishably increases to $F= 90\%$.  An important feature of the proposal is that the signal photon is an itinerant photon pulse, enabling detection of relatively wide-band  microwave photons.

\begin{figure}[Htb]
    \centering
\includegraphics[width = 3.2in]{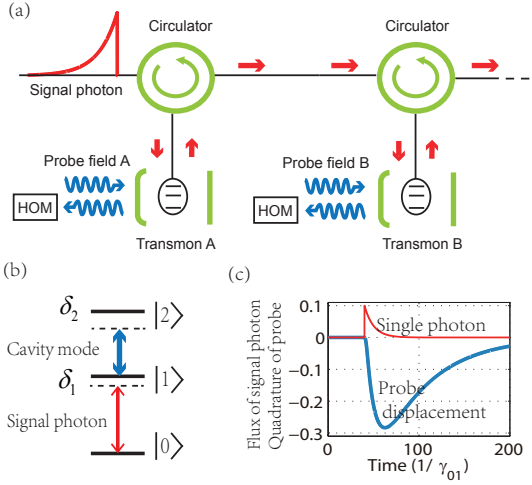}
\caption{(Color online) Schematic of  microwave photon counting. (a) A transmon qubit coupled to a microwave cavity provides the nonlinearity to detect the presence or absence of microwave photons propagating in the waveguide. If the waveguide is in the vacuum state, the transmon is transparent to the cavity field, so the probe field experiences no displacement; for a single signal photon, coherences in the transmon are produced, leading to a displacement of the probe field. 
A second transmon-cavity unit can be cascaded through a circulator to improve performance. (b) The energy level structure of the transmon. (c). The response of the cavity field to the incident single photon. }
\label{scheme}
\end{figure}

The scheme for single microwave photon detection is shown in Fig.\ \ref{scheme}. A transmon is embedded at one end of a waveguide, in which the signal (itinerant)
 microwave propagates.  The signal field is nearly resonant with the lowest transmon transition, $\ket{0}\leftrightarrow\ket{1}$. The transmon is also coupled to
 a coherently-driven microwave resonator, which is dispersively coupled with the second transmon transition.  The cavity is driven by an external coherent probe field, which ultimately yields information about the photon population in the signal field. This \emph{unit} (consisting of the transmon in a cavity) can be cascaded using circulators to achieve higher detection efficiency \cite{PhysRevLett.112.093601}. 

We first analyse a single unit, and later consider cascading several. In a rotating frame the Hamiltonian describing a unit is
\begin{eqnarray}
\hat{H}_s& =& \delta_1\hat{\sigma}_{11}+(\delta_1+\delta_2)\hat{\sigma}_{22}\nn
&-&ig_{12}(\hat{a}\hat{\sigma}_{21}-\hat{a}^\dag\hat{\sigma}_{12})-iE(\hat{a}-\hat{a}^\dag)
\end{eqnarray}
where $\hat{a}$ is the cavity annihilation operator, $g$ is the coupling strength between the cavity field and the transmon $\ket{1}\leftrightarrow\ket{2}$ transition, $E$ is the driving amplitude, and the detunings are  $\delta_1=\omega_{10}-\omega_s$, $\delta_2=\omega_{21}-\omega_{cav}$.

To model the itinerant signal  field, we invoke a fictitious source-cavity initially in a Fock state.  This field leaks out, producing an itinerant Fock state, which ultimately interacts with the transmon in the real cavity driven by the probe field.   The probe field reflected from the real cavity is measured by a homodyne detector. The resulting conditional system dynamics are described by the cascaded, stochastic master equation \cite{Car93,Gar93,gardiner2004quantum,wiseman2010quantum}:
\begin{eqnarray}
d\rho &=&dt\,\mathcal{L}\rho +dW\!(t)\,\mathcal{H}[e^{-i\phi}\sqrt{\kappa}\hat{a}]\rho,
\label{SME}
\end{eqnarray}
where
\begin{eqnarray}
 \mathcal{L}\rho&=&-i[\hat{H}_s,\rho]+\mathcal{D}[\sqrt{\gamma}_c\hat{c}]+\mathcal{D}[\sqrt{\gamma}_{01}\hat{\sigma}_{01}]\rho\nn
&+&\mathcal{D}[\sqrt{\gamma}_{12}\hat{\sigma}_{12}]\rho+\mathcal{D}[\sqrt{\kappa}\hat{a}]\rho \nn
&+& \sqrt{\gamma_c\gamma_{01}}([\hat{c}\rho,\hat{\sigma}_{10}]+[\hat{\sigma}_{01},\rho\hat{c}^\dag]),
\end{eqnarray}
and the corresponding Homodyne photocurrent is
\begin{eqnarray}
I(t)=\sqrt{\kappa}\av{e^{-i\phi}\hat{a}+e^{i\phi}\hat{a}^\dag}+dW(t)/dt
\end{eqnarray}
where $dW$ is a Weiner process satisfying $E[dW]=0$, $E[d^2W]=dt$, $\hat{c}$ is the annihilation operator of the source-cavity mode, $\gamma_c$ is the decay rate of the source-cavity (which determines the linewidth of the itinerant photon),  the phase angle $\phi$ is set by the local oscillator phase,
$
\mathcal{D}[\hat{r}]\rho =\frac{1}{2}(2\hat{r}\rho \hat{r}^{\dag
}-\rho \hat{r}^{\dag }\hat{r}-\hat{r}^{\dag }\hat{r}\rho )$, and $
\mathcal{H}[\hat{r}]\rho=\hat{r}\rho+\rho\hat{r}^\dag-{\rm Tr}[\hat{r}\rho+\rho\hat{r}^\dag]\rho
$.

Prior to arrival of the signal pulse, the cavity is driven by the probe field to its steady state, and  the transmon is initially in its ground state.  The itinerant signal photon pulse arrives at the transmon at time $t_0$.  Since the signal pulse decays over a finite time, the cavity field is transiently displaced from its steady state. This transient displacement is reflected in the homodyne photocurrent,
which thus contains information about the number of photons in the signal pulse.  There are several  methods to extract this information \cite{PhysRevA.76.012325}, the simplest of which is a linear filter applied to the homodyne current:
\begin{eqnarray}
S=\int^T_{t_0} I(t)h(t) dt,
\label{RF}
\end{eqnarray}
for some filter kernel $h$.  The optimal linear filter takes $h(t)=\bar I_1(t)$, where $\bar I_1$ is the expected homodyne current
 when there is a single signal photon. We have also implemented more sophisticated non-linear filters, using hypothesis testing \cite{PhysRevA.76.012325,tsang2012continuous},
which yields a small improvement, at a substantial computational cost.

As one measure of performance, we define a signal-to-noise ratio ${\snr}={(\bar{S_1}-\bar{S_0})}/{\sqrt{\textrm{Var}(S_1)+\textrm{Var}(S_0)}}$, where $S_n$ is the filter output conditioned on a signal pulse containing $n=0$ or 1 photons.
Due to the nonlinear interaction between the probe field and the transmon, $S_1$ is not a Gaussian variable, making $\snr$ difficult to interpret. Thus, we also report the distinguishability $F$, defined as the  probability of correctly inferring from the homodyne current the correct number of signal photons
\begin{equation}F=\big(P(S<S_{\rm th}|n=0)+P(S>S_{\rm th}|n=1)\big)/2,\label{F}\end{equation} where $S_{\rm th}$ is a threshold value for $S$ which discriminates
 between small and large probe displacement.  We have also assumed that $n=0,1$ are equally likely. 

\begin{figure}[Hbt]
    \centering
\includegraphics[width = \columnwidth]{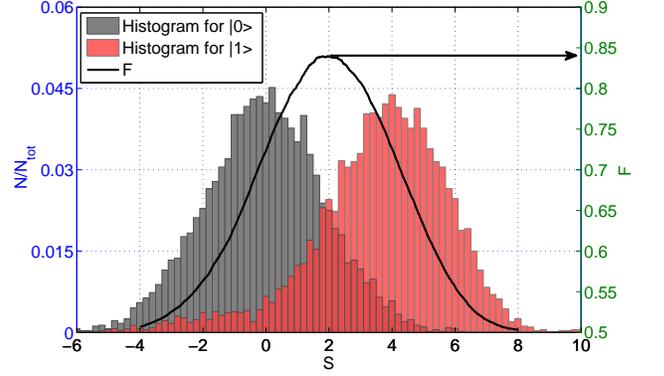}
\caption{(Color online) The histograms of filtered Homodyne signal for the presence/absence of the signal photon and the corresponding distinguishability. The black curve plots the distinguishability versus threshold values. The signal photon pulse is an exponentially-decayed pulse from a source cavity and the linear filter function is presented is Eq. (\ref{RF}). The parameters are: $\gamma_{01}=1$, $\gamma_{12}=0.1$, $g=2.45$, $\delta_{1}=-0.8$, $\delta_{2}=-18$, $\gamma_c=0.1$, $E=0.032$, $\kappa=0.037$, $\phi=\pi/2$, $t_0=0$ and $T=80$.}
\label{1tr}
\end{figure}%

To quantify the performance of a single unit as a photon detector, we perform a Monte-Carlo study, generating many trajectories with either $n=0$ or $n=1$,
 and computing $S$ for each.  Fig. \ref{1tr} shows  histograms of $S$ for $n=0$ (grey) and $n=1$ (red), for system parameters chosen to maximise $F$.
  The peaks of the histograms are reasonably distinguished.  The black trace shows $F$ as a function of $S_{\rm th}$.  We find $\snr_1=1.2$, and $F_1=84\%$  \footnote{The subscript denotes a single cavity-transmon unit.},
  which is a substantial improvement over \cite{PhysRevLett.110.053601}.  For comparison, the fidelity using the more sophisticated hypothesis testing filter
  gives slight improvement $F_1^{\textrm{HT}}=84.6\%$.

 We note that the optimal choice $\gamma_{12}=0.1\gamma_{01}$ used in Fig.\ \ref{1tr}  requires that  the microwave density-of-states (DoS) in the transmission line be engineered to suppress emission at $\omega_{12}$.  Without DoS engineering, $\gamma_{12}=2\gamma_{01}$ \cite{PhysRevA.76.012325}, and we find that the fidelity  is reduced to $F_1=81\%$.

The lifetime for the unit cavity is chosen to optimise  single-photon induced transmon excitation. Accordingly, the signal  pulse must be a relatively long, matching the  cavity life time. With a long pulse and a good cavity, during the interaction time of the signal photon with the system, the intra-cavity field changes dramatically (see Fig.\ 1c). In comparison, for situation without a unit cavity, the change in the probe is determined by the transmon coherence $\av{\hat{\sigma}_{12}}<\av{\hat{\sigma}_{11}}$, which decays quickly in that case. The  cavity allows the probe field to interact for a long time with the signal-induced coherence in the transmon, 
resulting in the larger integrated Homodyne signal over the measurement time.  


  The probe amplitude used in Fig.\ 2 was chosen to optimise the performance of the single-photon detector. Increasing the probe amplitude beyond this level leads to strong saturation effects in the transmon, consistent with the breakdown of an effective cross-Kerr description  as discussed in \cite{PhysRevLett.110.053601}.



The peak distinguishability for a single transmon is potentially useful in some applications.   To increase it further, we follow \cite{PhysRevLett.112.093601} and cascade multiple transmons using circulators to engineer a unidirectional waveguide. The computational cost of simulating a chain of transmons grows exponentially with the number of transmons, $N_{\rm tr}$, however it was shown in \cite{PhysRevLett.112.093601} that the SNR grows as $\sqrt{N_{\rm tr}}$, as might be expected for independent, repeated, noisy measurements of the same system. For our purposes, we consider cascading two transmons, A and B.  Since our detection process is non-absorbing, and circulators suppress back-scattering, the single microwave photon will deterministically interact with  A and then B in order, resulting in dynamical shifts for both cavity modes. We suppose that each cavity is addressed by a separate probe field, leading to two homodyne currents.  Again, we expect this to improve the SNR by $\sim\sqrt{2}$.

For computational efficiency in our Monte-Carlo simulations, we unravel the master equation to produce a stochastic Schrodinger equation \cite{wiseman2010quantum,PhysRevA.79.042103}, including four stochastic processes -- three quantum diffusion processes, one each for the cavity fields and an additional process to account for  cross-relaxation of the transmon into the waveguide,  and one quantum-jump process for the signal photon pulse \footnote{We emphasise that the jump process is solely to generate the homodyne currents (which are determined by the quantum diffusion processes); we do not subsequently use the jump records.}.
In the absence of a signal photon, the evolution of the unnormalized system wave function $\ket{\tilde{\psi}}$ is governed by
\begin{eqnarray}
d\ket{\tilde{\psi}(t)}&=&dt[-i(\hat{H}_s+\hat{H}_{cas})-\half(\sum_{j=A,B}\kappa_j\hat{a}^\dag_j\hat{a}_j+\hat{J}^\dag\hat{J}\nn
&+&\hat{J}^\dag_2\hat{J}_2)+\sum_{i=A,B}(e^{-i\phi_j}\sqrt{\kappa_j}\hat{a}_j) I^{(j)} + \hat{J}_2 I_2 ]\ket{\tilde{\psi}_c(t)},
\end{eqnarray}
where
\begin{eqnarray}
\hat{H}_s& =& \sum_{j=A,B}(\delta_{j1}\hat{\sigma}^{j}_{11}+(\delta_{j1}+\delta_{j2})\hat{\sigma}^{j}_{22}
-ig_j(\hat{a}_j\hat{\sigma}^{j}_{21}\nn
&-&\hat{a}^\dag_j\hat{\sigma}^{j}_{12})-iE_j(\hat{a}_j-\hat{a}^\dag_j))\nn
\hat{H}_{cas} &=& -\frac{i}{2}(\sum_{j=A,B}(\gamma_c\gamma^j_{01})^{1/2}\hat{c}\hat{\sigma}^j_{10}+({\gamma^A_{01}\gamma^B_{01}})^{1/2}\hat{\sigma}^A_{01}\hat{\sigma}^B_{10}\nn
&+&({\gamma^A_{12}\gamma^B_{12}})^{1/2}\hat{\sigma}^A_{12}\hat{\sigma}^B_{21})+h.c.
\end{eqnarray}
where we have defined     \mbox{$\hat{J}=\sqrt{}{\gamma_c}\hat{c}^\dag\hat{c}+\sum_{j=A,B}\sqrt{}{\gamma^j_{01}}\,\hat{\sigma}^j_{01}$} and
\mbox{$\hat{J}_2=\sum_{j=A,B}\sqrt{}{\gamma^j_{12}}\,\hat{\sigma}^j_{12}$}.
Upon a jump event in the signal field, the system state evolves discontinuously
\begin{eqnarray}
 \ket{\tilde{\psi}(t+dt)}=\hat{J}\ket{\tilde{\psi}(t)}.
\end{eqnarray}
The homodyne signals are given by $I^{(j)}, (j=A,B)$ for output of two cavities and $I_2$ for emission from transmons to the transmission line:
\begin{eqnarray}
I^{(j)}&=&\sqrt{\kappa_j}\av{e^{-i\phi_j}\hat{a}_j+e^{i\phi_j}\hat{a}_j}+dW_j/dt\nn
I_2&=& \av{\hat{J}_2+\hat{J}^\dag_2}+dW_{2}/dt
\end{eqnarray}

\begin{figure}[Hbt]
    \centering
\includegraphics[width = \columnwidth]{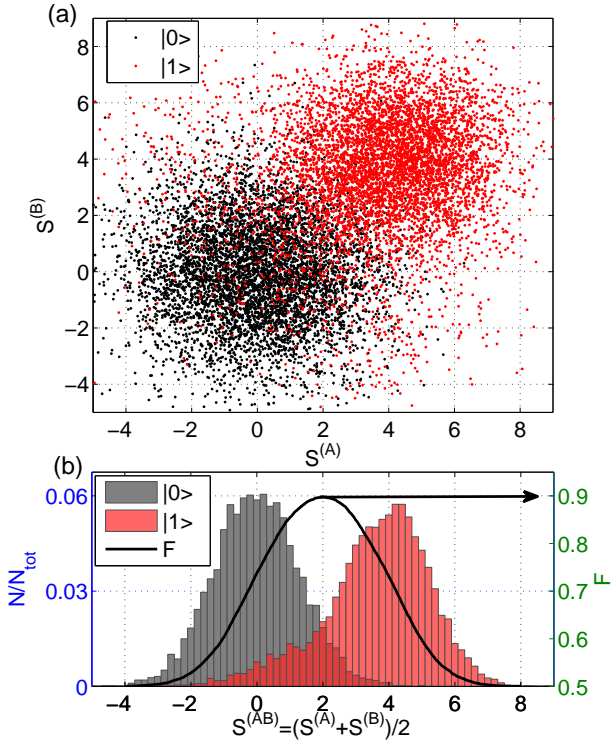}
\caption{ (Color online) (a) The scatter plot of the filtered Homodyne signals from two probe cavities with the presence/absence of the signal photon.  (b) The histogram of the sum Homodyne signal $S_{AB}$ and the corresponding distinguishability in the two cascaded transmons case. The parameters are:  $\gamma^A_{01}=\gamma^B_{01}=1$, $\gamma^A_{12}=\gamma^B_{12}=0.1$, $g_A=g_B=2.45$, $\delta_{1A}=\delta_{1B}=-0.8$, $\delta_{2A}=\delta_{2B}=-18$, $\gamma_c=0.1$, $E_A=E_B=0.032$, $\kappa_A=\kappa_B=0.037$, $\phi_A=\phi_B=\pi/2$, $t_0=0$ and $T=80$.}
\label{hist_2tr}
\end{figure}

We simulate 8000 trajectories using the same parameter values as before (assuming identical transmon-cavity units),  for each choice of $n$, to obtain a distribution of homodyne currents, $I^{(A)}$ and  $I^{(B)}$, which we integrate according to Eq.\ \ref{RF} to produce $S^{(A)}$ and $S^{(B)}$.  Fig.\ \ref{hist_2tr}(a) shows a scatter plot of the two homodyne signal pairs $(S^{(A)}_n,S^{(B)}_n)$ for $n=0$ (black) and $n=1$ (red). To  distinguish between these two distributions we project onto the sum  $S^{(AB)}=(S^{(A)}+S^{(B)})/2$, shown in Fig.\ \ref{hist_2tr}(b), and we calculate $\snr_2=1.7\approx\sqrt{2}\,\snr_1$, as expected.  Likewise, we define the distinguishability as in Eq.\ \ref{F}, replacing $S$ with  $S^{(AB)}$. Optimising $S_{\textrm{th}}$, we find $F=90\%$.  We note that if the distributions were in fact Gaussian, then this improvement in SNR would give a distinguishability of $91.5\%$, slightly higher than what we achieve.

\begin{figure}[Hbt]
    \centering
\includegraphics[width = 3.2in]{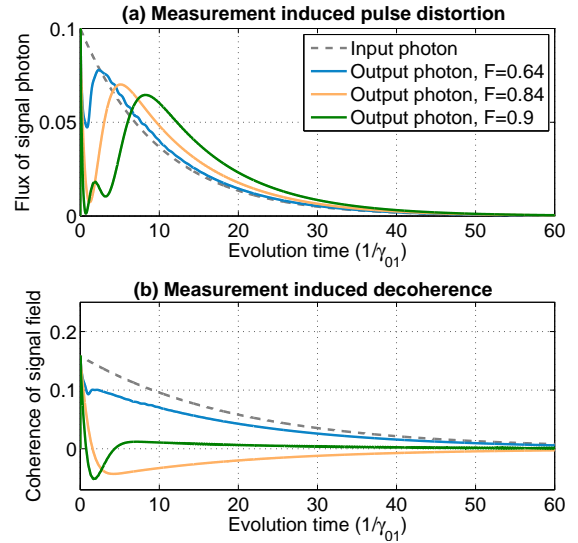}
\caption{ (Color online) (a) Pulse envelope distortion. (b) Measurement induced decoherence of the signal microwave photon state. The grey dash curves denote the input signal field and the solid curves denote the output signal field at different distinguishability. The blue ($\delta_2=-6$) and orange ($\delta_2=-18$) curves represent the output signal field after interacting with one transmon and the green ($\delta_{2A}=\delta_{2B}=-18$) curves represent the output signal field after interacting with two transmons. The other parameters are: $\gamma^A_{01}=\gamma^B_{01}=1$, $\gamma^A_{12}=\gamma^B_{12}=0.1$, $g_A=g_B=2.45$, $\delta_{1A}=\delta_{1B}=-0.8$, $\gamma_c=0.1$, $E_A=E_B=0.032$ and $\kappa_A=\kappa_B=0.037$.}
\label{flux_coh}
\end{figure}

In our proposed detector, there is in fact some distortion of the signal pulse envelope, as the transmon-cavity unit coherently
 interacts with the signal field, closely analogous to the pulse envelope distortion found in \cite{PhysRevLett.112.093601}.  This is shown in Fig.\ \ref{flux_coh}(a).  Here, we have allowed the detuning $\delta_2$ to vary,
  in order to  vary the distinguishability.  We see that the pulse envelope is maximally distorted  when $F$ is maximal, which follows
   since this is the condition under which the measurement back-action in maximised.
   For photon counting considered in this Letter, the deterministic pulse distortion is not a significant issue. However it may become so if the transmon were to be used to induce gates between photon-encoded states (e.g.\ in an interferometer), since the pulse-shape would encode some amount of `which-path' information leading to a reduction in coherence between different paths \cite{shapiro2006single}.  It may be possible to circumvent this problem, albeit at the cost of significant complexity \cite{chudzicki2013deterministic}.

Finally, we  consider what happens to a signal field that is prepared in a superposition of Fock states. In this case, QND measurement of the photon number should cause decoherence between  the components in the superposition, leaving populations unchanged
\cite{PhysRevLett.109.153601,PhysRevLett.77.4887}. Suppose $\hat{r}$ is an operator acting on the signal field. In a QND  number measurement, $[\hat{r}^\dag\hat{r},H_s]=0$,
 while $[\hat{r},H_s]\neq0$  so that the  coherence between Fock subspaces $\av{\hat{r}}$ decays during the interaction.
To demonstrate this effect, we take a superposition state $\ket{0}+\ket{1}$ as the initial state of the {fictitious source-cavity} and see how $\av{\hat{r}}$ evolves during the measurement process. Fig.\ \ref{flux_coh}(b) shows the time evolution of $\av{\hat{r}}$, 
 for different values of  distinguishability. This confirms that when the system is tuned to maximise the distinguishability, coherence is most rapidly suppressed. 

In summary, we have demonstrated a protocol for  photon counting of itinerant microwave photons,  which exploits the large cross-Kerr nonlinearity of a single transmon in a microwave waveguide \cite{PhysRevLett.111.053601}.  By synthesising results from \cite{PhysRevLett.110.053601,PhysRevLett.112.093601}, and adding a local cavity to each transmon, we find that we can cascade multiple such devices to produce  effective photon counters.  With just two, we achieve a distinguishability of $90\%$, which may be useful in certain microwave experiments.  We anticipate that 3 or 4 units could achieve fidelities up to $95\%$.


We thank Zhenglu Duan and Arkady Fedorov for helpful discussions. We acknowledge support from the China Scholarship Council and the Australian Research Council through grant CE110001013. JC was supported in part by National Science Foundation Grant Nos.~PHY-1212445 and~PHY-1314763, by Office of Naval Research Grant No.~N00014-11-1-0082. GJ acknowledges support from the Swedish Research Council and the EU through the ERC and the FP7 STREP PROMISCE.

\bibliography{bixuan}


\end{document}